# Enhancing the injection locking range of spin torque oscillators through mutual coupling

M. Romera [1]*, P. Talatchian [1], R. Lebrun [1], K. J. Merazzo [2,3], P. Bortolotti [1], L. Vila [2], J. D. Costa [4], R. Ferreira [4], P. P. Freitas [4], M.-C. Cyrille [2,3], U. Ebels [2], V. Cros [1], J. Grollier [1]

[1] Unité Mixte de Physique CNRS Thales, Univ. Paris-Sud, Université Paris-Saclay, 91767 Palaiseau, France
[2] Univ. Grenoble Alpes, CEA, CNRS, SPINTEC, F-38000 Grenoble, France
[3] Univ. Grenoble Alpes, CEA-LETI MINATEC, F-38000 Grenoble, France
[4] International Iberian Nanotechnology Laboratory (INL), 4715-31 Braga, Portugal
* miguel.romera@thalesgroup.com

We investigate how the ability of the vortex oscillation mode of a spin-torque nano-oscillator to lock to an external microwave signal is modified when it is coupled to another oscillator. We show experimentally that mutual electrical coupling can lead to locking range enhancements of a factor 1.64. Furthermore, we analyze the evolution of the locking range as a function of the coupling strength through experiments and numerical simulations. By uncovering the mechanisms at stake in the locking range enhancement, our results will be useful for designing spin-torque nano-oscillators arrays with high sensitivities to external microwave stimuli.

Spin-torque nano-oscillators (STNOs) are promising candidates for the next generation of multifunctional spintronic nano-devices [1] such as efficient integrated microwave generators [2], detectors [3,4] and bioinspired computing units [5]. One of their characteristic features compared to other auto-oscillators is their high non-linearity [6]. On one hand, this high non-linearity translates into one of their most attractive properties: their ability to easily adapt their frequencies to external stimuli. On the other hand, by increasing their sensitivity to noise, it has the undesirable effect of broadening the spectral linewidth. In the last few years, many efforts have been made to improve the spectral purity of spin-torque oscillators, a crucial step towards most applications. From these studies, a promising approach consists in the synchronization of the oscillators to an external microwave source [6-21] (injection locking), or to other oscillators [22-28] (mutual synchronization). In both cases, a crucial parameter to optimize for applications is the locking range. A first way to increase the locking range is to work in regimes where the oscillator's non-linearity is the largest. However, as mentioned earlier, this comes at the detriment of spectral purity. In the case of injection locking, another possibility is to increase the power of the external signal, but this is limited by the breakdown voltage of the tunnel barrier of the device and results in an increase of power consumption. Therefore, finding alternative paths to enhance the locking range is an important step towards designing next generation's magnetic microwave devices.

In this manuscript we combine experimental results with numerical simulations to show that the injection locking range of a spin-torque oscillator can be enhanced by coupling it to another spin-torque oscillator. Furthermore, it is shown that the locking range can be tuned by changing the mutual coupling strength between oscillators.

The experimental results are obtained for magnetic tunnel junctions with the following composition: Ta/CuN/Ta/PtMn(20)/CoFe(2)/Ru(0.85)/CoFeB(2.2)/CoFe(0.5)/MgO(1)/CoFeB(1.5)/Ta(0.2)/NiFe(7)/Ta. Here PtMn(20)/CoFe(2)/Ru(0.85)/CoFeB(2.2) is a synthetic ferrimagnet (SyF) uniformly magnetized in-plane, that is used as a polarizer, CoFeB(1.5)/Ta(0.2)/NiFe(7) is the free layer and numbers represent thickness in nanometers. Samples were grown by sputter-deposition and patterned down to the bottom electrode into circular nanopillars with a diameter of 200 nm. The nano-pillars exhibit a TMR of 64% and a resistance-area product of RA~1 $\Omega.\mu m^2$. With this combination of materials and geometry, the free layer contains a stable magnetic vortex as a ground state. A magnetic field of $H$=2.4 kOe is applied perpendicularly to the layers to get an efficient spin transfer torque acting on the vortex core [29]. dc current is injected perpendicularly to the layers to induce vortex dynamics, which leads to periodic oscillations of the magneto-resistance and translates into an oscillating voltage. Fig. 1a shows a schematic of the electrical setup. The two oscillators are connected in series, and electrically coupled via their own emitted microwave signals [28, 30]. dc current is supplied to the oscillators by two different sources, allowing an independent control of the current flowing through each oscillator, and therefore of their frequencies. A microwave source is used to inject an external microwave current of $P$=-15 dBm into the circuit, at frequencies around twice the carrier frequency of the oscillators. The total microwave signal from the two oscillators is recorded with a spectrum analyzer.

Using this configuration, we study how the injection locking of oscillator 1 to the external microwave source is modified by its coupling to oscillator 2. For this purpose, the current through oscillator 1 is kept fixed ($I_{STO1}$=6.3 mA), and we perform injection locking experiments for different values of the current applied through oscillator 2 ($I_{STO2}$). We focus our study on oscillator 1 because it is the oscillator which exhibits the largest ability to adapt its frequency to external stimuli. The curve with red filled squares in Fig. 1b shows the frequency of oscillator 1 as a function of the frequency of the source when oscillator 1 is uncoupled ($I_{STO2}$=0 mA). For frequencies of the source between 762.8 and 766.18 MHz, the frequency of oscillator 1 is locked to half the frequency of the source, resulting in a locking range of 1.69 MHz. We then couple the two oscillators together by sending a dc current through oscillator 2. The strength of coupling between oscillators is inversely proportional to their frequency difference [17] and can lead to mutual synchronization [28] when the frequency difference is small. Experimentally we have determined that this happens typically below ~2 MHz in these samples. In the present study we set the frequency of oscillator 2 close to the frequency of oscillator 1 so that the frequency difference is slightly larger than 2 MHz. Thus the oscillators do not synchronize but they are coupled and can influence each other.

The black filled dots in Fig. 1b correspond to the frequency of oscillator 1 when it is coupled to oscillator 2 (black open dots, $I_{STO2}$=3.25 mA). A clear single sided expansion of the injection locking range of oscillator 1 is observed compared to the case when oscillator 2 is not active (red squares in Fig. 1b; $I_{STO2}$=0 mA). The locking range grows from 1.69 MHz in the uncoupled case to 2.77 MHz in the coupled case.

Let's examine the mechanisms at the origin of this enhancement. When the frequency of the source is swept from left to right, at some point, oscillator 1 starts to get attracted by the source, and its frequency is pulled down towards half the frequency of the source. By decreasing, its frequency also gets closer to the frequency of oscillator 2. Both oscillators then interact more and more, and oscillator 2 starts to assist the source in pulling down the frequency of oscillator 1. Due to the additional force, the frequency of oscillator 1 decreases further and gets locked to the common frequency of the external source and oscillator 2. This happens at a frequency of the external signal ($F^{*loc}_{ext}$=760.6 MHz) well below the value at which oscillator 1 gets locked when it is uncoupled ($F^{loc}_{ext}$=762.8 MHz). In consequence, the injection locking range increases by 64% of its initial value.

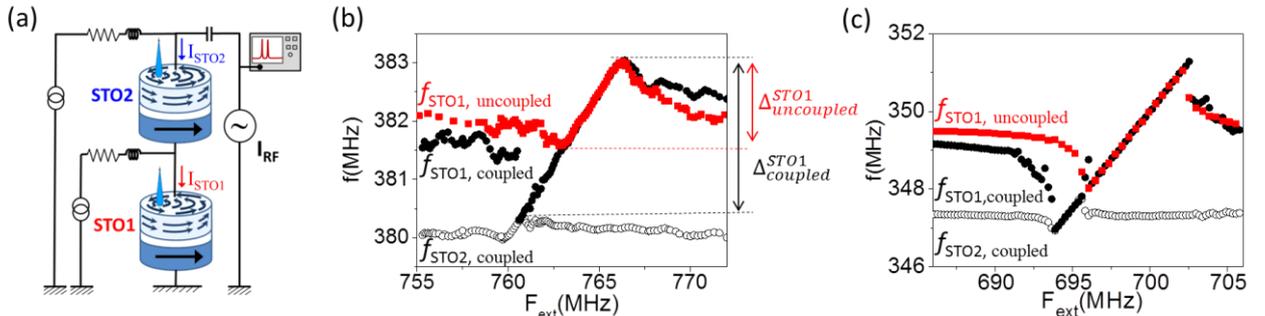

Figure 1: (a) Schematic of the electrical setup with two vortex oscillators electrically connected in series. Two different dc current sources are used to have an independent control of the current flowing through each oscillator. (b) Injection locking experiments at P= -15 dBm with oscillator 1 uncoupled ($I_{STO2}$= 0 mA, red squares) and with oscillator 1 coupled to oscillator 2 ($I_{STO2}$= 3.25 mA, black dots; solid dots represent the frequency of oscillator 1 and open dots the one of oscillator 2). The current applied to oscillator 1 is kept fixed ($I_{STO1}$= 6.3 mA). Vertical arrows highlight the injection locking range of oscillator 1, delimited by horizontal dashed lines. (c) Numerical simulations of injection locking with oscillator 1 uncoupled ($I_{STO2}$= 0 mA, red squares) and coupled to oscillator 2

($I_{STO2}$= 3.57 mA, black dots; solid dots represent the frequency of oscillator 1 and open dots the one of oscillator 2). The current applied to oscillator 1 is kept fixed ($I_{STO1}$= 2.6 mA).

In order to confirm these assumptions, we have performed numerical simulations that are shown in Fig. 1c, in the uncoupled case (red squares) and in the coupled case (black dots). In the simulations, the magnetization dynamics of two electrically coupled vortex oscillators is obtained by solving numerically the differential Thiele equation [31,32]:

$$G_i \times \frac{dX_i}{dt} - \widehat{D}_i(X_i)\frac{dX_i}{dt} - \frac{\partial W_i(X_i, I_{STOi}, I_{com}^{rf})}{\partial X_i} + F_i^{STT}(X_i, I_{STOi}, I_{com}^{rf}) = 0 \quad (1)$$

simultaneously for the two vortex $i$ =1,2. Here, $X_i = \begin{pmatrix} x_i \\ y_i \end{pmatrix}$ is the vortex core position, $G_i$ is the gyrovector, $\widehat{D}_i$ is the damping, $W_i$ is the potential energy of the vortex, $F_i^{STT}$ is the spin-transfer force and $I_{STOi}$ is the dc current applied to oscillator $i$. The total microwave current $I_{com}^{rf}$ flowing through the oscillators consists of the external microwave current provided by the source, as well as the microwave currents emitted by the oscillators themselves (see Supplementary Material for details) [33].

The material parameters considered are extracted from the analytical fitting of the experimental response of each oscillator and are given in the Supplementary Material. Similarly to the experimental approach, the dc current through oscillator 1 is kept constant along the whole study ($I_{STO1}$=2.6 mA). The frequency of each oscillator is extracted from the $5\mu s$ time trace of the angular evolution of the vortex core trajectories. The simulations point out the field like torque as responsible for the coupling mechanism leading to injection locking at half the frequency of the source (see section D in the Suppl. Mat.).

As can be seen in Fig. 1c, the injection locking range of oscillator 1 increases from 3 MHz to 4.32 MHz due to its coupling to oscillator 2, corresponding to an enhancement of 42%. These numerical results are in good agreement with the behavior observed experimentally (Fig. 1b). In particular, both in experiments and simulations the locking range's enhancement is unidirectional in the sense that it is the left boundary of the locking range that changes while the right boundary remains constant. This is because oscillator 2 is the responsible of this enhancement and has a lower frequency than oscillator 1.

In order to evaluate the dependence of the locking range of oscillator 1 on the strength of its coupling to oscillator 2, we have varied $I_{STO2}$ while keeping $I_{STO1}$ constant. Under these conditions an increase of $I_{STO2}$ translates into a decrease of the frequency detuning between the oscillators and consequently to an increase of the coupling strength.

Injection locking experiments performed for different values of $I_{STO2}$ and the corresponding numerical simulations are shown in Figs. 2(a-e) and 3(a-e) respectively. In both cases, the first graph on the left (Figs. 2a and 3a) corresponds to oscillator 1 uncoupled ($I_{STO2}$=0 mA). Figs. 2(b-e) and 3(b-e) correspond to injection locking with the two oscillators coupled, for values of $I_{STO2}$ increasing from left (Figs. 2b and 3b) to right (Figs. 2e and 3e).

Extracting the locking range for the different dc current values, we display in Fig. 2f how the locking range of oscillator 1 is enhanced when it is coupled to oscillator 2, with respect to the reference value of oscillator 1 uncoupled ($\frac{\Delta_{coupled}^{STO1} - \Delta_{uncoupled}^{STO1}}{\Delta_{uncoupled}^{STO1}} \cdot 100$), as a function of $I_{STO2}$. A striking result is that the locking range of oscillator 1 can be enhanced up to a value 65% larger than the reference locking range by taking advantage of the coupling to oscillator 2. Furthermore, the locking range enhancement can be tuned by controlling the coupling strength between oscillators through $I_{STO2}$. Fig. 2f shows three different trends in the evolution of the locking range with $I_{STO2}$. In good agreement, the simulations (Fig. 3f) show the same qualitative bell shape behavior with three regions in the locking range's dependence with $I_{STO2}$.

The following regions can be distinguished:

- Region 1 (R1 in Fig. 2f) corresponds to $I_{STO2}$ lower than 3 mA. In this region, the frequency detuning between the oscillators is large leading to weak influence of the coupling. The mechanism of enhancement of oscillator 1's locking range is the frequency pulling provoked by the coupling to oscillator 2 when their frequencies get closer. As an example see Fig. 2b. Here, when the frequency of oscillator 2 increases due to injection locking and gets close to the frequency of oscillator 1, this one is strongly pulled down. As a consequence the locking range of oscillator 1 is enhanced up to 2.56 MHz, corresponding to an enhancement of 53% with respect to the value of oscillator 1 uncoupled. The numerical simulations reproduce well this behavior as can be seen in Figs. 3b and 3c. An additional feature occurs in Figs. 3b and 3c, just before oscillator 1 gets locked to the source. A linear response of its frequency with a different slope than the one given by the source is observed. This effect, which is more evident in the simulations but is also present experimentally, has been identified as the locking of oscillator 1 to a modulation signal given by $F_{ext}$ - $f_{STO2}$.

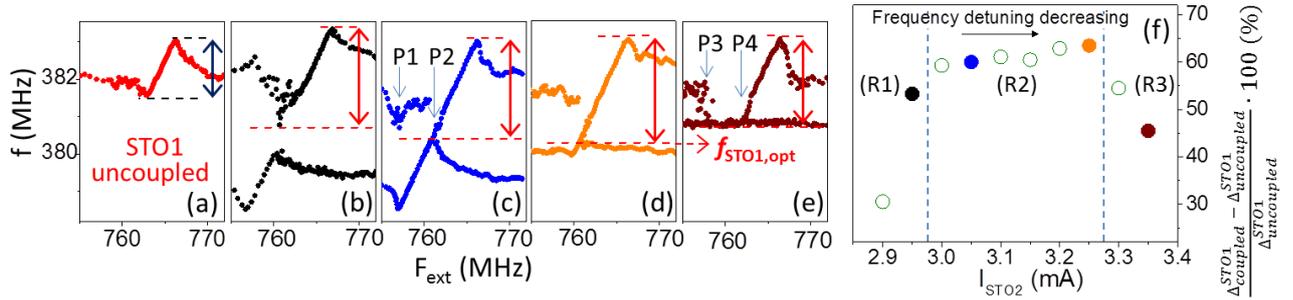

Figure 2: (a-e) Injection locking experiments at 2f and P= -15 dBm of oscillator 1 uncoupled (a) or coupled to oscillator 2 (b-e) for different values of the current flowing through oscillator 2. Arrows highlight the injection locking range of oscillator 1, delimited by horizontal dashed lines. (f) Enhancement of the injection locking range of oscillator 1 due to its coupling to oscillator 2 ($\frac{\Delta^{STO1}_{coupled}-\Delta^{STO1}_{uncoupled}}{\Delta^{STO1}_{uncoupled}} \cdot 100$) as a function of the current applied on oscillator 2. The frequency difference between oscillators decreases as $I_{STO2}$ increases. Filled dots are from the data in panels (a) to (e), having the same color.

In region 1, the injection locking range of oscillator 1 increases sharply upon increasing $I_{STO2}$ (see Fig. 2f). This is because increasing $I_{STO2}$ has two consequences, both contributing to the increase of locking range in this region. (i) The frequency detuning between the oscillators is reduced, and oscillator 2 locks better and better to the source. Thus the frequency of oscillator 2 increases through frequency locking, interacting more with oscillator 1, and pulling it down more efficiently. (ii) The power emitted by oscillator 2 increases, which translates into an increase of its ability to attract oscillator 1.

- Region 2 (R2 in Fig. 2f) corresponds to $I_{STO2}$ between 3 mA and 3.25 mA. In this region, the frequency detuning is smaller than in region 1, and the coupling is stronger. Here, the enhancement of oscillator 1's injection locking range occurs through locking oscillator 1 simultaneously to oscillator 2 and the external source. Fig. 2c shows an example with two interesting features. First, a frequency pulling effect is observed in oscillator 1 towards the frequency of oscillator 2 (point P1 in Fig. 2c). The frequency of oscillator 1 is here already well below the minimum frequency of the reference case of oscillator 1 uncoupled (Fig. 2a). Upon increasing $F_{ext}$, oscillator 1 gets eventually locked to the signal emitted by both the external source and oscillator 2. This happens when the frequency of oscillator 2 increases due to frequency locking to the source so that both oscillator's frequencies get very close (point P2 in Fig. 2c). The locking range is in this case 2.74 MHz, meaning an enhancement of 62% with respect to oscillator 1 uncoupled. A similar behavior is observed when the frequency detuning is reduced further (Fig. 2d): oscillator 1 gets locked to oscillator 2 at values of $F_{ext}$ around the end of the locking range of oscillator 2. This behavior is well reproduced by the simulations, as can be seen in Fig. 3d.

In region 2, the injection locking range of oscillator 1 increases upon increasing $I_{STO2}$ in a slower manner than in region 1 (Fig. 2f). This is because the reduction of frequency detuning upon increasing $I_{STO2}$ does not affect the locking range of oscillator 1 in this region. The reason is that oscillator 1 actually gets locked (not only frequency pulled) to the common frequency of oscillator 2 and the external source. In addition, in the region of $F_{ext}$ where oscillator 1 gets locked, oscillator 2 is already locked to the source, so its frequency is $F_{ext}/2$. Therefore, even though the running frequency of oscillator 2 depends on $I_{STO2}$, there is a range of values of $F_{ext}$ for which the frequency of oscillator 2 is the same ($f_{STO2}=F_{ext}/2$) independently of its running frequency. The frequency locking of both oscillators happens in this range of $F_{ext}$ when their frequencies are closer than a certain threshold, or in other words, when the frequency of oscillator 2 reaches a certain value ($F_{ext}^{thr}/2$). Oscillator 1 gets locked to this frequency value $F_{ext}^{thr}/2$, which sets the beginning of its locking range (see section F in the Suppl. Mat. for a schematic). $F_{ext}^{thr}/2$ is independent of the frequency detuning. Thus the locking range of oscillator 1 is not affected by the frequency detuning in region 2.

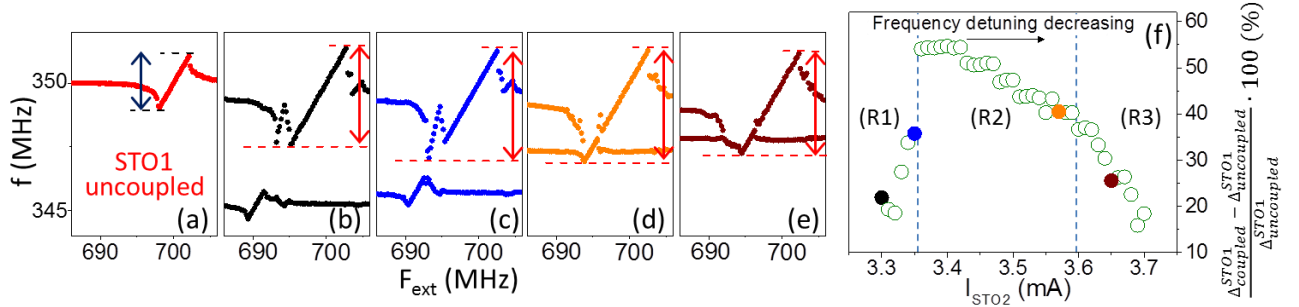

Figure 3: (a-e) Numerical simulations of the injection locking at 2f of oscillator 1 uncoupled (a) or coupled to oscillator 2 (b-e) for different values of the current flowing through oscillator 2. Arrows highlight the injection locking range of oscillator 1, delimited by horizontal dashed lines. (f) Enhancement of the simulated injection locking range of oscillator 1 due to its coupling to oscillator 2 ($\frac{\Delta_{coupled}^{STO1} - \Delta_{uncoupled}^{STO1}}{\Delta_{uncoupled}^{STO1}} \cdot 100$) as a function of the current applied on oscillator 2. The frequency difference decreases as $I_{STO2}$ increases. Filled dots are from the data in panels (a) to (e), having the same color.

On the other hand, upon increasing $I_{STO2}$, the power emitted by oscillator 2 increases, which increases the coupling strength. Thus oscillator 2 can attract oscillator 1 for larger frequency detuning [9]. This is the reason why oscillator 1's locking range increases slightly with $I_{STO2}$ in region 2. Optimum conditions are obtained for $I_{STO2}$=3.25 mA, with oscillator 1 being locked to frequencies as low as $f_{STO1,opt}$=380.31 MHz and the locking range being enhanced by 64%.

- Finally, region 3 (R3 in Fig. 2f) corresponds to $I_{STO2}$ higher than 3.25 mA. Here, the frequency detuning is smaller than in region 2, and the coupling strength is larger. From the analysis of the microwave properties of oscillator 2 when it is uncoupled, we know that this region of large $I_{STO2}$ is associated to large power and low non-linearity of oscillator 2 (see sections B and C in the Suppl. Mat.). Thus (i) the locking range of oscillator 2 is very small, and (ii) oscillator 2 strongly attracts oscillator 1. An example is shown in Fig. 2e. As can be seen, the frequency pulling of oscillator 1 towards the external signal leads first to the frequency locking of oscillator 1 to oscillator 2, which is not yet locked to the external source (point P3 in Fig. 2e). Upon increasing $F_{ext}$, oscillator 1 gets eventually locked to the source when $F_{ext}/2$ is larger than the frequency of oscillator 2 (point P4, Fig. 2e). This behavior is again well reproduced by numerical simulations (Fig. 3e).

It is noted that the left boundary of the locking range is considered to be the minimum frequency reached by oscillator 1. This corresponds to the frequency of oscillator 2 in Fig. 2e. As shown in Fig. 2f, the locking range of oscillator 1 decreases upon increasing $I_{STO2}$ in region R3. The main reason is that the frequency of oscillator 2 is already above the frequency at which oscillator 1 gets locked at optimum conditions ($f_{STO1,opt}$ reached for $I_{STO2}$=3.25 mA, see Fig. 2d). Furthermore, due to the small injection locking range of oscillator 2 in this region, the frequency of oscillator 2 remains above $f_{STO1,opt}$ along the whole injection locking experiment (see Fig. 2e). Thus, oscillator 1 gets locked to the frequency of oscillator 2 above $f_{STO1,opt}$. Upon increasing $I_{STO2}$, the frequency of oscillator 2, which is the minimum frequency reached by oscillator 1, increases and the injection locking range of oscillator 1 decreases.

To summarize, we have shown that electrical coupling between spin-torque oscillators can be used to enhance the injection locking range of the oscillators to an external source. Interestingly, the locking range can be tuned by controlling the coupling strength. In our approach based on electrical coupling this can be easily done by controlling the dc current flowing through each oscillator. In particular, the locking range of one oscillator can be tuned by controlling the dc current flowing through the other oscillator. We have shown that the mechanism responsible for the locking range enhancement can be either frequency pulling or frequency locking between oscillators, which lead to different dependences of the locking range on the frequency detuning. These results will be important for microwave applications where injection locking to an external source is an issue such as frequency emitters. They also have the potential to open new paths towards neuromorphic computing with spin-torque nano-oscillators, where inputs are microwave stimuli, and outputs are the different synchronization states of coupled oscillators arrays [5,34-38].

Supplementary Material: (a) Details on the numerical simulations and material parameters, (b) fvsI and ΔfvsI curves for the free running oscillators, (c) injection locking range of oscillator 2 coupled and uncoupled as a function of $I_{STO2}$, (d) simulated injection locking with and without field like torque, (e) contour plot of an injection locking experiment, and (f) schematic of the injection locking of two coupled oscillators with the conditions of region R2.

This work was supported by the French National Research Agency (ANR) under Contract ANR-14-CE26-0021 (MEMOS), by the French RENATECH network, by ON2 project INTEGRATION (grant NORTE-07-0124-FEDER-000050), by the EC under the FP7 program No. 317950 MOSAIC and by the ERC bioSPINspired 682955.